
\magnification=1200\overfullrule=0pt\baselineskip=15pt
\vsize=22truecm \hsize=15truecm \overfullrule=0pt\pageno=0

\font\titlefont=cmbx10 scaled \magstep1
\font\sectnfont=cmbx8  scaled \magstep1
\def\mname{\ifcase\month\or January \or February \or March \or April
           \or May \or June \or July \or August \or September
           \or October \or November \or December \fi}
\def\date{\hbox{\strut\mname \number\year}}
\def\banner{\hfill\hbox{\vbox{\offinterlineskip\crnum}}\relax}
\def\manner{\hbox{\vbox{\offinterlineskip\crnum\date}}
               \hfill\relax}
\footline={\ifnum\pageno=0\manner\else\hfil\number\pageno\hfil\fi}
%
%
%
\newcount\FIGURENUMBER\FIGURENUMBER=0
\def\FIG#1{\expandafter\ifx\csname FG#1\endcsname\relax
               \global\advance\FIGURENUMBER by 1
               \expandafter\xdef\csname FG#1\endcsname
                              {\the\FIGURENUMBER}\fi}
\def\figtag#1{\expandafter\ifx\csname FG#1\endcsname\relax
               \global\advance\FIGURENUMBER by 1
               \expandafter\xdef\csname FG#1\endcsname
                              {\the\FIGURENUMBER}\fi
              \csname FG#1\endcsname\relax}
\def\fig#1{\expandafter\ifx\csname FG#1\endcsname\relax
               \global\advance\FIGURENUMBER by 1
               \expandafter\xdef\csname FG#1\endcsname
                      {\the\FIGURENUMBER}\fi
           Fig.~\csname FG#1\endcsname\relax}
\def\figand#1#2{\expandafter\ifx\csname FG#1\endcsname\relax
               \global\advance\FIGURENUMBER by 1
               \expandafter\xdef\csname FG#1\endcsname
                      {\the\FIGURENUMBER}\fi
           \expandafter\ifx\csname FG#2\endcsname\relax
               \global\advance\FIGURENUMBER by 1
               \expandafter\xdef\csname FG#2\endcsname
                      {\the\FIGURENUMBER}\fi
           figures \csname FG#1\endcsname\ and
                   \csname FG#2\endcsname\relax}
\def\figto#1#2{\expandafter\ifx\csname FG#1\endcsname\relax
               \global\advance\FIGURENUMBER by 1
               \expandafter\xdef\csname FG#1\endcsname
                      {\the\FIGURENUMBER}\fi
           \expandafter\ifx\csname FG#2\endcsname\relax
               \global\advance\FIGURENUMBER by 1
               \expandafter\xdef\csname FG#2\endcsname
                      {\the\FIGURENUMBER}\fi
           figures \csname FG#1\endcsname--\csname FG#2\endcsname\relax}
\newcount\TABLENUMBER\TABLENUMBER=0
\def\TABLE#1{\expandafter\ifx\csname TB#1\endcsname\relax
               \global\advance\TABLENUMBER by 1
               \expandafter\xdef\csname TB#1\endcsname
                          {\the\TABLENUMBER}\fi}
\def\tabletag#1{\expandafter\ifx\csname TB#1\endcsname\relax
               \global\advance\TABLENUMBER by 1
               \expandafter\xdef\csname TB#1\endcsname
                          {\the\TABLENUMBER}\fi
             \csname TB#1\endcsname\relax}
\def\table#1{\expandafter\ifx\csname TB#1\endcsname\relax
               \global\advance\TABLENUMBER by 1
               \expandafter\xdef\csname TB#1\endcsname{\the\TABLENUMBER}\fi
             Table \csname TB#1\endcsname\relax}
\def\tableand#1#2{\expandafter\ifx\csname TB#1\endcsname\relax
               \global\advance\TABLENUMBER by 1
               \expandafter\xdef\csname TB#1\endcsname{\the\TABLENUMBER}\fi
             \expandafter\ifx\csname TB#2\endcsname\relax
               \global\advance\TABLENUMBER by 1
               \expandafter\xdef\csname TB#2\endcsname{\the\TABLENUMBER}\fi
             Tables \csname TB#1\endcsname{} and
                    \csname TB#2\endcsname\relax}
\def\tableto#1#2{\expandafter\ifx\csname TB#1\endcsname\relax
               \global\advance\TABLENUMBER by 1
               \expandafter\xdef\csname TB#1\endcsname{\the\TABLENUMBER}\fi
             \expandafter\ifx\csname TB#2\endcsname\relax
               \global\advance\TABLENUMBER by 1
               \expandafter\xdef\csname TB#2\endcsname{\the\TABLENUMBER}\fi
            Tables \csname TB#1\endcsname--\csname TB#2\endcsname\relax}
\newcount\REFERENCENUMBER\REFERENCENUMBER=0
\def\REF#1{\expandafter\ifx\csname RF#1\endcsname\relax
               \global\advance\REFERENCENUMBER by 1
               \expandafter\xdef\csname RF#1\endcsname
                         {\the\REFERENCENUMBER}\fi}
\def\reftag#1{\expandafter\ifx\csname RF#1\endcsname\relax
               \global\advance\REFERENCENUMBER by 1
               \expandafter\xdef\csname RF#1\endcsname
                      {\the\REFERENCENUMBER}\fi
             \csname RF#1\endcsname\relax}
\def\ref#1{\expandafter\ifx\csname RF#1\endcsname\relax
               \global\advance\REFERENCENUMBER by 1
               \expandafter\xdef\csname RF#1\endcsname
                      {\the\REFERENCENUMBER}\fi
             [\csname RF#1\endcsname]\relax}
\def\refto#1#2{\expandafter\ifx\csname RF#1\endcsname\relax
               \global\advance\REFERENCENUMBER by 1
               \expandafter\xdef\csname RF#1\endcsname
                      {\the\REFERENCENUMBER}\fi
           \expandafter\ifx\csname RF#2\endcsname\relax
               \global\advance\REFERENCENUMBER by 1
               \expandafter\xdef\csname RF#2\endcsname
                      {\the\REFERENCENUMBER}\fi
             [\csname RF#1\endcsname--\csname RF#2\endcsname]\relax}
\def\refand#1#2{\expandafter\ifx\csname RF#1\endcsname\relax
               \global\advance\REFERENCENUMBER by 1
               \expandafter\xdef\csname RF#1\endcsname
                      {\the\REFERENCENUMBER}\fi
           \expandafter\ifx\csname RF#2\endcsname\relax
               \global\advance\REFERENCENUMBER by 1
               \expandafter\xdef\csname RF#2\endcsname
                      {\the\REFERENCENUMBER}\fi
            [\csname RF#1\endcsname,\csname RF#2\endcsname]\relax}
\def\refs#1#2{\expandafter\ifx\csname RF#1\endcsname\relax
               \global\advance\REFERENCENUMBER by 1
               \expandafter\xdef\csname RF#1\endcsname
                      {\the\REFERENCENUMBER}\fi
           \expandafter\ifx\csname RF#2\endcsname\relax
               \global\advance\REFERENCENUMBER by 1
               \expandafter\xdef\csname RF#2\endcsname
                      {\the\REFERENCENUMBER}\fi
            [\csname RF#1\endcsname,\csname RF#2\endcsname]\relax}
\def\refss#1#2#3{\expandafter\ifx\csname RF#1\endcsname\relax
               \global\advance\REFERENCENUMBER by 1
               \expandafter\xdef\csname RF#1\endcsname
                      {\the\REFERENCENUMBER}\fi
           \expandafter\ifx\csname RF#2\endcsname\relax
               \global\advance\REFERENCENUMBER by 1
               \expandafter\xdef\csname RF#2\endcsname
                      {\the\REFERENCENUMBER}\fi
           \expandafter\ifx\csname RF#3\endcsname\relax
               \global\advance\REFERENCENUMBER by 1
               \expandafter\xdef\csname RF#3\endcsname
                      {\the\REFERENCENUMBER}\fi
     [\csname RF#1\endcsname,\csname RF#2\endcsname,\csname
              RF#3\endcsname]\relax}
\def\Ref#1{\expandafter\ifx\csname RF#1\endcsname\relax
               \global\advance\REFERENCENUMBER by 1
               \expandafter\xdef\csname RF#1\endcsname
                      {\the\REFERENCENUMBER}\fi
             Ref.~\csname RF#1\endcsname\relax}
\def\Refs#1#2{\expandafter\ifx\csname RF#1\endcsname\relax
               \global\advance\REFERENCENUMBER by 1
               \expandafter\xdef\csname RF#1\endcsname
                      {\the\REFERENCENUMBER}\fi
           \expandafter\ifx\csname RF#2\endcsname\relax
               \global\advance\REFERENCENUMBER by 1
               \expandafter\xdef\csname RF#2\endcsname
                      {\the\REFERENCENUMBER}\fi
        Refs.~\csname RF#1\endcsname{},\csname RF#2\endcsname\relax}
\def\Refto#1#2{\expandafter\ifx\csname RF#1\endcsname\relax
               \global\advance\REFERENCENUMBER by 1
               \expandafter\xdef\csname RF#1\endcsname
                      {\the\REFERENCENUMBER}\fi
           \expandafter\ifx\csname RF#2\endcsname\relax
               \global\advance\REFERENCENUMBER by 1
               \expandafter\xdef\csname RF#2\endcsname
                      {\the\REFERENCENUMBER}\fi
            Refs.~\csname RF#1\endcsname--\csname RF#2\endcsname]\relax}
\def\Refand#1#2{\expandafter\ifx\csname RF#1\endcsname\relax
               \global\advance\REFERENCENUMBER by 1
               \expandafter\xdef\csname RF#1\endcsname
                      {\the\REFERENCENUMBER}\fi
           \expandafter\ifx\csname RF#2\endcsname\relax
               \global\advance\REFERENCENUMBER by 1
               \expandafter\xdef\csname RF#2\endcsname
                      {\the\REFERENCENUMBER}\fi
        Refs.~\csname RF#1\endcsname{} and \csname RF#2\endcsname\relax}
\newcount\EQUATIONNUMBER\EQUATIONNUMBER=0
\def\EQ#1{\expandafter\ifx\csname EQ#1\endcsname\relax
               \global\advance\EQUATIONNUMBER by 1
               \expandafter\xdef\csname EQ#1\endcsname
                          {\the\EQUATIONNUMBER}\fi}
\def\eqtag#1{\expandafter\ifx\csname EQ#1\endcsname\relax
               \global\advance\EQUATIONNUMBER by 1
               \expandafter\xdef\csname EQ#1\endcsname
                      {\the\EQUATIONNUMBER}\fi
            \csname EQ#1\endcsname\relax}
\def\EQNO#1{\expandafter\ifx\csname EQ#1\endcsname\relax
               \global\advance\EQUATIONNUMBER by 1
               \expandafter\xdef\csname EQ#1\endcsname
                      {\the\EQUATIONNUMBER}\fi
            \eqno(\csname EQ#1\endcsname)\relax}
\def\EQNM#1{\expandafter\ifx\csname EQ#1\endcsname\relax
               \global\advance\EQUATIONNUMBER by 1
               \expandafter\xdef\csname EQ#1\endcsname
                      {\the\EQUATIONNUMBER}\fi
            (\csname EQ#1\endcsname)\relax}
\def\eq#1{\expandafter\ifx\csname EQ#1\endcsname\relax
               \global\advance\EQUATIONNUMBER by 1
               \expandafter\xdef\csname EQ#1\endcsname
                      {\the\EQUATIONNUMBER}\fi
          Eq.~(\csname EQ#1\endcsname)\relax}
\def\eqand#1#2{\expandafter\ifx\csname EQ#1\endcsname\relax
               \global\advance\EQUATIONNUMBER by 1
               \expandafter\xdef\csname EQ#1\endcsname
                        {\the\EQUATIONNUMBER}\fi
          \expandafter\ifx\csname EQ#2\endcsname\relax
               \global\advance\EQUATIONNUMBER by 1
               \expandafter\xdef\csname EQ#2\endcsname
                      {\the\EQUATIONNUMBER}\fi
         Eqs.~\csname EQ#1\endcsname{} and \csname EQ#2\endcsname\relax}
\def\eqto#1#2{\expandafter\ifx\csname EQ#1\endcsname\relax
               \global\advance\EQUATIONNUMBER by 1
               \expandafter\xdef\csname EQ#1\endcsname
                      {\the\EQUATIONNUMBER}\fi
          \expandafter\ifx\csname EQ#2\endcsname\relax
               \global\advance\EQUATIONNUMBER by 1
               \expandafter\xdef\csname EQ#2\endcsname
                      {\the\EQUATIONNUMBER}\fi
          Eqs.~\csname EQ#1\endcsname--\csname EQ#2\endcsname\relax}
%
\newcount\SECTIONNUMBER\SECTIONNUMBER=0
\newcount\SUBSECTIONNUMBER\SUBSECTIONNUMBER=0
\def\section#1{\global\advance\SECTIONNUMBER by 1\SUBSECTIONNUMBER=0
      \bigskip\goodbreak\line{{\sectnfont \the\SECTIONNUMBER.\ #1}\hfil}
      \smallskip}
\def\subsection#1{\global\advance\SUBSECTIONNUMBER by 1
      \bigskip\goodbreak\line{{\sectnfont
         \the\SECTIONNUMBER.\the\SUBSECTIONNUMBER.\ #1}\hfil}
      \smallskip}
%
%
\def\DT{\Delta\tau}\def\dt{\ifmmode\DT\else$\DT$\fi}
\def\BETAC{\beta_c}\def\betac{\ifmmode\BETAC\else$\BETAC$\fi}
\def\LARGE{8\times16^3}\def\large{\ifmmode\LARGE\else$\LARGE$\fi}
\def\SMALL{8\times12^3}\def\small{\ifmmode\SMALL\else$\SMALL$\fi}
\def\MQ{ma}\def\mq{\ifmmode\MQ\else$\MQ$\fi}
\def\REL{Re(L)}\def\rel{\ifmmode\REL\else$\REL$\fi}
\def\ABL{Abs(L)}\def\absl{\ifmmode\ABL\else$\ABL$\fi}
\def\PPBAR{\overline\chi\chi}\def\ppbar{\ifmmode\PPBAR\else$\PPBAR$\fi}
\def\ord{\ifmmode\langle\REL\rangle\else$\langle\REL\rangle$\fi}
\def\cord{\ifmmode\langle\PPBAR\rangle\else$\langle\PPBAR\rangle$\fi}
\def\pprgi{\ifmmode\langle\PPBAR\rangle_{RGI}\else$\langle\PPBAR
           \rangle_{RGI}$\fi}
\def\entropy{\ifmmode{s\over T^3}
             \else$s/T^3$\fi}
\def\NT{N_\tau}\def\nt{\ifmmode\NT\else$\NT$\fi}
\def\NS{N_\sigma}\def\ns{\ifmmode\NS\else$\NS$\fi}
\def\GT{g_\tau}\def\gt{\ifmmode\GT\else$\GT$\fi}
\def\GS{g_\sigma}\def\gs{\ifmmode\GS\else$\GS$\fi}
\def\CT{c'_\tau}\def\ct{\ifmmode\CT\else$\CT$\fi}
\def\CS{c'_\sigma}\def\cs{\ifmmode\CS\else$\CS$\fi}
\def\NF{n_\f}\def\nf{\ifmmode\NF\else$\NF$\fi}
\def\TC{T_c}\def\tc{\ifmmode\TC\else$\TC$\fi}

\def\lsim{\raise0.3ex\hbox{$<$\kern-0.75em\raise-1.1ex\hbox{$\sim$}}}

\def\gsim{\raise0.3ex\hbox{$>$\kern-0.75em\raise-1.1ex\hbox{$\sim$}}}
%

\def\f{{\scriptscriptstyle f}}
%

%
%
%
%
%
%
\magnification=1200\overfullrule=0pt\baselineskip=15pt
\hoffset=0.3truecm
\voffset=1.51truecm
\vsize=22truecm \hsize=15truecm \overfullrule=0pt\pageno=0

\font\titlefont=cmbx10 scaled \magstep2
\font\daggerfont=cmmi10 scaled \magstep3
\font\sectnfont=cmbx10 scaled \magstephalf
\def\mname{\ifcase\month\or January \or February \or March \or April
           \or May \or June \or July \or August \or September
           \or October \or November \or December \fi}
\def\date{\hbox{\strut\mname \number\year}}
\def\binum{\hbox{BI-TP 95/29\strut}}
\def\banner{\hfill\hbox{\vbox{\offinterlineskip\binum\date}}\relax}
\def\manner{\hbox{\vbox{\offinterlineskip}}
               \hfill\relax}
\def\lsim{\raise0.3ex\hbox{$<$\kern-0.75em\raise-1.1ex\hbox{$\sim$}}}
\footline={\ifnum\pageno=0\manner\else\hfil\number\pageno\hfil\fi}

%
\def\ZP{{ Z.\ Phys.\ }}
\def\PR{{ Phys.\ Rev.\ }}

\def\PRL{{ Phys.\ Rev.\ Lett.\ }}
\def\PL{{ Phys.\ Lett.\ }}
\def\NP{{ Nucl.\ Phys.\ }}
\def\OP{{\langle L \rangle }}
\def\OPQ{{\langle L^2 \rangle }}
\def\MO{{\langle |L| \rangle }}
%
%
%
%
\banner\bigskip\begingroup\titlefont\obeylines
\vskip 20pt
\centerline{Critical behaviour of SU(2) lattice gauge theory}
\centerline{A complete analysis with the ${\bf \chi^2}$-method
$^{^{\daggerfont\dag}}$}
\vskip 10pt
\endgroup\bigskip
\footnote{}{$^a$Fakult\"at f\"ur Physik, Universit\"at Bielefeld,
 D-33615 Bielefeld, Germany}
\footnote{}{$^b$N.N.~Bogolyubov Institute for Theoretical Physics,
252143 Kiev, Ukraine}
\footnote{}{$^{\dagger}$Work supported by NATO, Linkage grant No. 930224,
and the Deutsche}
\footnote{}{$^{ }$ Forschungsgemeinschaft, grant Pe 340/3-3}
\bigskip
\centerline{J.~Engels$^a$, S.~Mashkevich$^b$, T.~Scheideler$^a$
 and G.~Zinovjev$^b$}
\bigskip
\bigskip
\bigskip
\bigskip
\centerline{{\bf
ABSTRACT}}\medskip We determine the critical point and the ratios
$\beta/\nu$ and  $\gamma/\nu$ of critical exponents of the
deconfinement transition in $SU(2)$ gauge theory
by applying the $\chi^2$-method to Monte Carlo data of the
modulus and the square of the Polyakov loop. With the same technique
we find from the Binder cumulant $g_r$ its universal value at the
critical point in the thermodynamical limit to $-1.403(16)$ and for
the next-to-leading exponent $\omega=1\pm0.1$. From the derivatives
of the Polyakov loop dependent quantities we estimate then $1/\nu$.
The result from the derivative of $g_r$ is $1/\nu=0.63\pm0.01$,
in complete agreement with that of the $3d$ Ising model.
\vfil\eject
\centerline{\bf 1. Introduction }
\medskip
 The investigation of the critical properties of any physical theory
on a finite lattice is a non-trivial problem.
This is so, because the singularity of the infinite volume
theory at a phase transition is modified due to the finite volume
into a rapid but nonsingular variation of the order parameter.
The location of the critical point becomes then unclear and the
infinite volume form of observables with critical behaviour
like the order parameter, the susceptibility and higher cumulants
cannot be used for the determination of the critical exponents.
The observed finite size effects are on the other hand controlled
by these exponents, as described by finite size scaling (FSS) theory.
Simulations on lattices of varying size
allow then to calculate the critical parameters of the theory.

Recently, a new method, the $\chi^2$-method, was proposed
\ref{Eng0} and applied to the
expectation value of the square of the magnetization.
Though this quantity is not peaked at the transition, it was shown
that one may deduce from its FSS behaviour the precise asymptotic
critical point and the ratio of the critical exponents
 $\gamma$ and $\nu$ in a very straightforward manner.

In this letter we want to extend the method to the modulus
and Binder's fourth-order cumulant $g_r$ \ref{Bind} of the magnetization.
The intention is to find the ratio $\beta/\nu$
of exponents and to check the consistency of
the results for the critical point from the different variables.
As it turns out it is possible to obtain still further information
on universal quantities and by examining the derivatives of the
variables at the critical point one may even determine the inverse of
the exponent $\nu$ of the correlation length with - at least for
$SU(2)$ - unreached precision.
\vskip 15pt
\centerline{\bf 2. Finite size scaling theory and the $\chi^2$-method}
\medskip
 Let us briefly review that part of FSS theory which is relevant
for our considerations. In the neighbourhood of the critical
temperature $T_c$ of a second order phase transition
one expects in the limit of large $N_\sigma$, the characteristic
length scale of the system, for the magnetization or order parameter
(denoted by $\OP$ here) that
$$
\OP \sim (T - T_c)^\beta ,~{\rm for}~~ T\rightarrow T_c^+ .
\EQNO{infl}
 $$
 The behaviour near to $T_c$ of the susceptibility
$\chi$ and the correlation length $\xi$ in the large $\NS$ limit
is expected to be
$$
\chi  \sim |T - T_c|^{- \gamma} ,
\EQNO{infc}
$$
$$
{\xi} \sim |T - T_c|^{- \nu} .
\EQNO{infx}
$$
A quantitative analysis of the finite lattice modifications to these
formulae becomes possible by using renormalization
group theory. In this framework it has been shown \ref{Barb} that
the singular part of the free energy density has the following form
$$
f_s(x,h,N_\sigma ) = N_\sigma^{-d}Q_{f_s}(g_T N_\sigma^{1/\nu},
g_h N_\sigma^{{(\beta + \gamma )}/ \nu}, g_i N_\sigma^{y_i})~.
\EQNO{free}
$$
The scaling function $Q_{f_s}$ depends on the temperature $T$ and the
external field strength $h$ in terms of a thermal and a magnetic
scaling field
$$
g_T = c_T x + O(xh,x^2)~,
\EQNO{gt}
$$
$$
g_h = c_h h + O(xh,h^2)~,
\EQNO{gh}
$$
which are independent of $\NS$ and where $x$ is the reduced temperature
$$
x = {{T - T_{c, \infty}} \over {T_{c, \infty}}}~.
\EQNO{redt}
$$

Here the action contains a further symmetry breaking term
$h \cdot N_{\sigma}^d L$. Also
additional irrelevant scaling fields $g_i$ with
negative exponents $y_i$ may be present.
\medskip
The order parameter $\OP$, the susceptibility $\chi$
and the cumulant $g_r$ are obtained from $f_s$ by taking
derivatives with respect to $h$ at $h=0$.
 The general form of the scaling relations derived in this way is
$$
O(x,N_\sigma) = N_{\sigma}^{\rho / \nu} \cdot Q_O
(g_T N_{\sigma}^{1/{\nu}},g_i N_{\sigma}^{y_i})~.
\EQNO{scale}
$$
Here $O$ is $\OP, \chi$ and $g_r$ with
$(\rho = - \beta, \gamma~{\rm and}~ 0.)$
Taking into account only the largest irrelevant exponent $y_1=-\omega$ and
expanding the scaling function $Q_O$ to first order at $x = 0$
results in the following equation
$$
O (x,N_\sigma) =  N_{\sigma}^{\rho/ \nu}
[ c_0 + (c_1 + c_2 N_{\sigma}^{-\omega}) x
N_{\sigma}^{1/ \nu} + c_3 N_{\sigma}^{-\omega} ]~.
\EQNO{expan}
$$
Standard FSS methods are based on the evaluation
of \eq{scale} and/or \eq{expan}
in the neighbourhood of the infinite volume critical coupling.
\medskip
The $\chi^2$-method \ref{Eng0} utilizes \eq{expan} only at $x=0$
$$
O (x=0,N_\sigma) =  N_{\sigma}^{\rho/ \nu}
[ c_0 + c_3 N_{\sigma}^{-\omega} ]~,
\EQNO{expan0}
$$
and determines the critical point
 from the leading $\NS$-behaviour of \eq{expan0}.
This is motivated by the fact, that for $x \not= 0$ the
$\NS$-behaviour is drastically changed due to the presence of
$x\NS^{1/\nu}$-terms ( for $SU(2)~{\rm e.g.}~1/\nu \approx 1.6 $ \ref{Land}).
The critical point is then defined as that point where a fit to
the leading $\NS$-behaviour of \eq{expan0} has the least minimal $\chi^2$.
\medskip
We have two different forms of fits.
If the exponent $\rho \not= 0$ the leading $\NS$-behaviour is
given only by the first term in \eq{expan0}.
Taking then the logarithm we find
$$
\ln O = \ln c_0 +{\rho \over \nu} \ln N_{\sigma}~~,
\EQNO{lnex}
$$
i.e. we have a linear dependence on ln$\NS$ with slope $\rho/\nu$.
Linear fits to $\ln O$ as a function of $\ln \NS$ give then also the
exponent ratio. A different form of fit is to be used in the case
$\rho=0$, where the leading $\NS$-behaviour is
$$
O  = c_0 + c_3 N_{\sigma}^{-\omega}~.
\EQNO{direx}
$$
\medskip
It is interesting to consider also
derivatives of the observables with respect to $x$. Since the scaling
functions $Q_O$ are depending on $x\NS^{1/\nu}$ we obtain
$$
{\partial O \over \partial x}(x,N_\sigma) = N_{\sigma}^{(1+\rho)/\nu}
\cdot Q_O^{\prime}(xN_{\sigma}^{1/{\nu}})~,
\EQNO{deriv}
$$
where the prime denotes the derivative with respect to the argument;
the dependence on the irrelevant fields is not explicitly shown, but
present. Since the derivative of the scaling function is again a
scaling function, derivatives of observables $O$ may be used to find
the value of $(1+\rho)/\nu$ from fits to the leading $\NS$-behaviour of
$$
{\partial O \over \partial x} (x=0,N_\sigma) =  N_{\sigma}^{(1+\rho)/ \nu}
[ c_1 + c_2 N_{\sigma}^{-\omega} ]~.
\EQNO{expan0d}
$$
Derivatives may also be used to define new direct scaling functions by
$$
U = x{ \partial \ln O \over \partial x } = U (x\NS^{1/\nu})~.
\EQNO{ufun}
$$
If for any finite value of $\NS$ the scaling function $Q_O$ and its
derivative $Q_O^{\prime}$ are finite in a neighbourhood of $x=0$ --
which is the normal case -- we have immediately
$$
U(0) = 0.
\EQNO{ucrit}
$$
In the thermodynamic limit, i.e. for $\NS \rightarrow \infty$, we
get, however a different result. From \eqto{infl}{infx}, or in general
 from
$$
O_{\infty} = c_O x^{-\rho} {\rm~~~for~~~}x\rightarrow +0~,
\EQNO{ocrit}
$$
we find
$$
U_{\infty}(0) = -\rho~.
\EQNO{ucrit}
$$
An attempt to determine $\rho$ at $x=0$ from infinite volume formulae
will therefore usually fail on finite lattices. Yet at small, but finite,
positive $x$ and large $\NS$, i.e. for large arguments $xN_{\sigma}^{1/{\nu}}$,
$U$ will approach $-\rho$.

\vskip 15pt
\centerline{\bf 3. SU(2) gauge theory at finite temperature}
\medskip
\indent Let us consider $SU(2)$ gauge theory on
 $N_\sigma^3\times N_\tau$
lattices using the standard Wilson action
$$
S(U) =  {4\over g^2} \sum_p (1 -{1\over 2}Tr U_p)~,
\EQNO{action}
$$
where $U_p$ is the product of link operators around a plaquette. The
number of lattice points in the space (time) direction
$N_{\sigma(\tau)}$
and the lattice spacing $a$ fix the volume and temperature as
$$
V = (N_\sigma a)^3 ,~~~~ T = 1 / (N_\tau a)~.
\EQNO{volte}
$$
\indent On an infinite volume lattice the order parameter or
magnetization for the
deconfinement transition is the expectation value of the Polyakov loop
$$
L({\bf x}) =  {1\over 2} Tr \prod_{\tau=1}^{N_\tau}U_{\tau,
{\bf x};4}~,
\EQNO{polya}
$$
or else, that of its lattice average
$$
L = {1\over N_{\sigma}^3}\sum_{\bf x}L({\bf x})~,
\EQNO{latta}
$$
where $U_{x;4}$ are the $SU(2)$ link matrices at four-position $x$
in time direction.
\indent Since, due to system flips between the two ordered states on
finite lattices the expectation value $\OP$ is always zero,~
the true susceptibility
$$
\chi = N_\sigma^3(\OPQ - \OP^2)~,
\EQNO{susc}
$$
reduces there to
$$
\chi_v = N_\sigma^3\OPQ~.
\EQNO{suscv}
$$
The quantity $\chi_v$ is monotonically rising as a function of the
coupling $4/g^2$ or the temperature $T$. Below the critical point
$$
\chi_v = \chi ~,
\EQNO{renco}
$$
and one expects therefore $\chi_v$ to have the same FSS behaviour as the true
susceptibility for $T \lsim T_c$. In ref.~\ref{Eng0} it was shown
that this is indeed the case.

In the following we want to extend the analysis of ref.~\ref{Eng0} to the
remaining magnetization dependent observables and their derivatives.
In particular, we want to take the expectation value of the modulus
of the lattice average, $\MO$, as a kind of ``finite lattice
order parameter", hoping as usual \refs{Land}{Eng1} that
its FSS behaviour is controlled by
the exponent $\beta$ of the infinite volume order parameter.
In addition we want to investigate the behaviour of the normalized
fourth cumulant of the magnetization
$$
g_r = { {\langle L^4 \rangle} \over {\OPQ}^2 } - 3~,
\EQNO{gr1}
$$
which is directly a scaling function and corresponds, up to a constant,
to the renormalized coupling defined for infinite systems \ref{Tous}.

\vskip 15pt
\centerline{\bf 4. Numerical results and discussion}
\medskip


The $SU(2)$ Monte Carlo data, which we want to use here
were computed on $\NS^3 \times \NT$
lattices with $\NS = 8,12,18,26$ and $\NT = 4$ and have already
been reported on in refs. \refs{Eng1}{Eng2} and \ref{Eng0}.
Since in the latter reference the critical point was determined
to $4/g^2_{c,\infty}=2.2988(1)$, we calculated in addition the still
missing results at this point for $\NS=8,12$ and 18 and
$\NT=4$ with high statistics. They are presented in \table{MCdata}.
Proceeding in the same way as in ref. \ref{Eng0}, we evaluated
then the data in the very close vicinity
( $2.298 \leq 4/g^2 \leq 2.300$)
of the transition with the density of states method (DSM) \ref{Dens}.
The newly calculated MC data were in complete agreement with the
former DSM interpolation and introduced only a tiny shift to the
interpolation itself. As usual, the errors were determined with
the Jackknife method.
\medskip

$$\vbox{\offinterlineskip
\halign{
\strut\vrule     \hfil $#$ \hfil  &
      \vrule # & \hfil $#$ \hfil  &
      \vrule # & \hfil $#$ \hfil  &
      \vrule # & \hfil $#$ \hfil  &
      \vrule # & \hfil $#$ \hfil
      \vrule \cr
\noalign{\hrule}
  ~N_{\sigma }~
&&~N_{\tau }~
&&~{\MO}~
&&~{\OPQ}~
&&~{g_r}~\cr
\noalign{\hrule}
 ~~~~8~~~&&~~~4~~~&&~~0.14343(18)~~&&~~0.025723(54)~~&&~~-1.3532(20)~~\cr
 ~~~12~~~&&~~~4~~~&&~~0.11591(22)~~&&~~0.016749(48)~~&&~~-1.3695(37)~~\cr
 ~~~18~~~&&~~~4~~~&&~~0.09370(88)~~&&~~0.01091(15)~~~&&~~-1.386(16)~~~\cr
 ~~~26~~~&&~~~4~~~&&~~0.07649(75)~~&&~~0.00733(10)~~~&&~~-1.365(17)~~~\cr
\noalign{\hrule}}
}$$
\vskip 5pt
\centerline {\bf \table{MCdata}}
\vskip 5pt
\vbox{\baselineskip=10pt
\noindent
\centerline{MC results at $4/g^2=2.2988$.}
}\baselineskip=15pt
\medskip
\noindent
At each $4/g^2$-value we have made linear $\chi^2$-fits to our
DSM interpolation results. For $\MO$ and $\chi_v$ the form of \eq{lnex}
was used, i.e. ln$\MO$ and ln$\chi_v$ were fitted as a function of
ln$\NS$. Correspondingly, \eq{direx} was utilized for fixed values
of $\omega$ near to 1 to perform linear fits of $g_r$ as a function
of $\NS^{-\omega}$. In Fig. 1 we compare the obtained minimal
$\chi^2/N_f$ for the three quantities.
Here the number of degrees of freedom, $N_f$, is 2, and $\omega$ was
fixed to 1.

We observe, that all three observables give consistent results in
several respects : the positions of the minima of the
$\chi^2_{min}/N_f$-curves are nearly coinciding -- their differences give
us an error estimate on the critical coupling; the curves have the same
shape and show a very sharp increase, symmetric around the respective
minimum; the size of $\chi^2_{min}/N_f$ close to the minimum is of
order one, which indicates that our error estimates are reasonable.
Finally, the minima themselves are very close to zero, which means
that the finite size behaviours of the observables at the critical
point are indeed described by the formulae, \eqand{lnex}{direx}.
In \table{mincoup}, we list the minimum positions found.

\medskip

$$\vbox{\offinterlineskip
\halign{
\strut\vrule     \hfil $#$ \hfil  &
      \vrule # & \hfil $#$ \hfil  &
      \vrule # & \hfil $#$ \hfil
      \vrule \cr
\noalign{\hrule}
  ~~O~
&&~~4/g^2_{min}~
&&~~{\omega}~\cr
\noalign{\hrule}
     ~~\MO~&&~~2.29895~~&&~~~~~~~~~~~~\cr
\noalign{\hrule}
 ~~~~\chi_v~~~&&~~2.29890~~&&~~~~~~~~~~~~\cr
\noalign{\hrule}
 ~~~~g_r~~~&&~~2.29905~~&&~~0.9,~1.0~~\cr
  ~~~~~~~~~&&~~2.29900~~&&~~~~~1.1~~~~\cr
\noalign{\hrule}}
}$$
\vskip 5pt
\centerline {\bf \table{mincoup}}
\vskip 5pt
\vbox{\baselineskip=10pt
\noindent
\centerline{Minimum positions of $\chi^2_{min}/N_f$.}
}\baselineskip=15pt
\medskip
\noindent
We conclude from their variations, that the critical point is at
$$
4/g^2_{c,\infty}=2.29895(10)~.
\EQNO{critp}
$$
Inside error bars, this value is compatible with the one of ref.
\ref{Eng0}, which was determined from the finite volume susceptibility
$\chi_v$ only. The shift in $4/g^2_{min}(\chi_v)$ due to the increased
statistics is 0.0001.
As can be seen from \table{mincoup} a variation of $\omega$ in the
range 0.9-1.1 has nearly no effect on the position of the minimum.
Increasing $\omega$ further shifts the minimum point to smaller
$4/g^2$-values, the width of the $\chi^2_{min}/N_f$-curve becomes
smaller, but at the same time the minimim value increases slightly.
Decreasing $\omega$ leads to opposite effects: shift of
$4/g^2_{min}$ to higher values, widening of the $\chi^2_{min}/N_f$-curve,
decrease of the minimum. A value of $\omega=1.0$ or slightly higher
seems reasonable, because then the width of the $\chi^2_{min}/N_f$-curve
is the same as that for $\MO$. Also the minimum position is then in
good agreement with that of the other observables. In
ref. \ref{Eng2}, where the critical points for $\NT=4$ and 6 were
determined from the intersection points of the $g_r$-curves for
different $\NS$, a value of~$\omega=0.9$ was found for $\NT=6$,
whereas for $\NT=4$ the contribution of irrelevant scaling fields
was below visibility. With our approach we confirm the existence of
these fields also for $\NT=4$ with an exponent $\omega$ of comparable
size. In addition, $\omega \approx 1.0$ is the value found in the
$3d$ Ising model \ref{Land}, which belongs to the same universality
class as 3+1 dimensional $SU(2)$ gauge theory \ref{Svet}.

Each of our three fits yields one universal quantity: $\beta/\nu$ and
$\gamma/\nu$ are obtained from the $\MO$ and $\chi_v$ fits; the
constant $c_0$ from the $g_r$ fit is the universal value $g_r^{\infty}$
of $g_r$ at the critical point for $\NS\rightarrow\infty$. Since all
universal quantities should be equal for theories of the same
universality class, we compare our results to those of the $3d$ Ising
model \ref{Land}. This is shown in Fig. 2. Obviously we find excellent
agreement between the two theories. In \table{results} we have listed
the corresponding numbers at $4/g^2_{c,\infty}$. Their errors come from
two sources:
the uncertainty in $4/g^2_{c,\infty}$ and the fit errors at fixed $4/g^2$.
The value of $g_r^{\infty}$ was estimated in ref. \ref{Eng2}
 from Binder's \ref{Bind} "cumulant crossing" method to -1.38(5).
Our fit leads to a lower result -1.403(16), which is in closer agreement
with the Ising value -1.41. This comes about, because we include a
correction to scaling term in the fit ($c_3\NS^{-\omega}$), where
$c_3$ turns out to be positive.

As a final application of FSS methods we have investigated the derivatives
of our observables with respect to the coupling $4/g^2$. Since the reduced
temperature $x$ may be approximated by
$$
x = {{4/g^2 - 4/g^2_{c, \infty}} \over {4/g^2_{c, \infty}}}~,\EQNO{xdef}
$$
\noindent
we have proportionality between $\partial O /\partial x$ and
$\partial O /\partial (4/g^2)$. As a consequence \eq{expan0d} may be
used to find $(1+\rho)/\nu$ from a fit of the type of \eq{lnex} at the
critical point.
 \medskip

$$\vbox{\offinterlineskip
\halign{
\strut\vrule     \hfil $#$ \hfil  &
      \vrule # & \hfil $#$ \hfil  &
      \vrule # & \hfil $#$ \hfil  &
      \vrule # & \hfil $#$ \hfil
      \vrule \cr
\noalign{\hrule}
  ~{\rm Source}~
&&~~
&&~{SU(2)}~
&&~{\rm Ising~\ref{Land}}~\cr
\noalign{\hrule}
  ~~\MO~~&&~~\beta/\nu~~&&~~0.525(8)~~~&&~~0.518(7)~~~~\cr
  ~D\MO~~&&~~(1-\beta)/\nu~~&&~~1.085(14)~~&&~~1.072(7)~~~~\cr
 ~~~~~~~~~&&~~1/\nu~~&&~~1.610(16)~~&&~~1.590(2)~~~~\cr
 ~~~~~~~~~&&~~\nu~~&&~~0.621(6)~~~&&~~0.6289(8)~~\cr
 ~~~~~~~~~&&~~\beta~~&&~~0.326(8)~~~&&~~~0.3258(44)~\cr
\noalign{\hrule}
 ~~~~\chi_v~~~&&~~\gamma/\nu~~&&~~1.944(13)~~&&~~1.970(11)~~~\cr
 ~~~D\chi_v~~~&&~~(1+\gamma)/\nu~~&&~~3.555(15)~~&&~~3.560(11)~~~\cr
 ~~~~~~~~~&&~~1/\nu~~&&~~1.611(20)~~&&~~1.590(2)~~~~\cr
 ~~~~~~~~~&&~~\nu~~&&~~0.621(8)~~~&&~~0.6289(8)~~\cr
 ~~~~~~~~~&&~~\gamma~~&&~~1.207(24)~~&&~~1.239(7)~~~~\cr
\noalign{\hrule}
 ~~~~~~~~~&&~~\gamma/\nu+2\beta/\nu~~&&~~2.994(21)~~&&~~3.006(18)~~\cr
\noalign{\hrule}
 ~~~~g_r~~&&~~-g_r^{\infty}~~&&~~1.403(16)~~&&~1.41~~~~~~~\cr
 ~~~Dg_r~~&&~~1/\nu~~&&~~1.587(27)~~&&~~1.590(2)~~\cr
 ~~(\omega=1)~~&&~~\nu~~&&~~0.630(11)~~&&~~0.6289(8)~\cr
\noalign{\hrule}}
}$$
\vskip 5pt
\centerline {\bf \table{results}}
\vskip 5pt
\vbox{\baselineskip=10pt
\noindent
\centerline{Results from the $\chi^2$-method at
$4/g^2_{c,\infty}=2.29895(10)$.}
\centerline{Here, $DO$ denotes the derivative $\partial O /\partial (4/g^2)$.}
}\baselineskip=15pt
\medskip
\noindent
Whereas it is relatively easy to calculate the derivatives
 from spline interpolations to the DSM results, it seems more difficult to
estimate their errors. To achieve this, we applied the Jackknife method
to those spline derivatives, which were obtained from the individual
Jackknife block results. In Fig. 3 we show the slope parameters from the
derivative fits in the vicinity of the critical point. Again we find a
surprising agreement with the $3d$ Ising model. In \table{results} the
corresponding numbers are given as well as the resulting $1/\nu (\nu)$
and exponent values. At the critical point the minimal $\chi^2/N_f$
was of the order $3-4$ for the derivatives of $\MO$ and $\chi_v$ and
about 1 for the derivative of $g_r$. We think therefore, that the result
$$
\nu = 0.630(11)~.
\EQNO{nugr}
$$
\noindent
 From the derivative of $g_r$ is the more reliable one, though all
results are compatible with each other.
We have checked also the hyperscaling relation
$$
\gamma/\nu + 2\beta/\nu = d~.
\EQNO{hyper}
$$
As can be seen from \table{results} our results are perfectly
consistent with \eq{hyper}.

In summarizing the following comments are in order. We have shown,
that also for the more complex (as compared to the Ising model)
$SU(2)$ gauge theory it is possible to determine the universal
quantities related to the deconfinement transition with high
precision from simulations on finite lattices. The special FSS
method which we applied here, the $\chi^2$-method, proved to be
simple and accurate and showed consistency in the determination
of the critical point from different observables. An indispensible
part of the method is the DSM interpolation of the MC data.
The quality of the DSM interpolation was usually superior to any
single direct MC calculation apart from those with extremely high
statistics.

{\centerline{\bf References}}
\vskip 15pt

\item{\reftag{Eng0})}
J.~Engels, J.~Fingberg and V.~K.~Mitrjushkin, \PL B298 (1993) 154.
\item{\reftag{Bind})}
K.~Binder, \ZP B43 (1981) 119.
\item{\reftag{Barb})}
M.N. Barber, in: Phase Transitions and Critical
Phenomena, Vol. 8, eds. C. Domb and J.L. Lebovitz, Academic,
New York, 1983, p. 146.
\item{\reftag{Land})}
A.M.~Ferrenberg and D.P.~Landau, \PR B44 (1991) 5081.
\item{\reftag{Eng1})}
J.~Engels, J.~Fingberg and M.~Weber, \NP B332 (1990) 737.
\item{\reftag{Tous})}
M.N.~Barber, R.B.~Pearson, D.~Toussaint and J.L.~Richardson,
\PR B32 (1985) 1720.
\item{\reftag{Eng2})}
J.~Engels, J.~Fingberg and D.E.~Miller, \NP B387 (1992) 501.
\item{\reftag{Dens})}
G.~Bhanot, S.~Black, P.~Carter and R.~Salvador, \PL 183B (1986) 331;\break
G.~Bhanot, K.~Bitar, S.~Black, P.~Carter and R.~Salvador,
 \PL 187B (1987) 381;
G.~Bhanot, K.~Bitar and R.~Salvador, \PL 188B (1987) 246;
\item{}
M.~Falconi, E.~Marinari, M.L.~Paciello, G.~Parisi
 and B.~Taglienti, \PL 108B (1982) 331;
E.~Marinari, \NP B235 (1984) 123;
\item{}
A.M.~Ferrenberg and R.H.~Swendsen, \PRL 61 (1988) 2635
and \PRL 63 (1989) 1195.
\item{\reftag{Svet})}
B.~Svetitsky and G.~Yaffe, \NP B210 [FS6] (1982) 423.

\vfil\eject

\centerline{\bf Figure Captions}

\bigskip

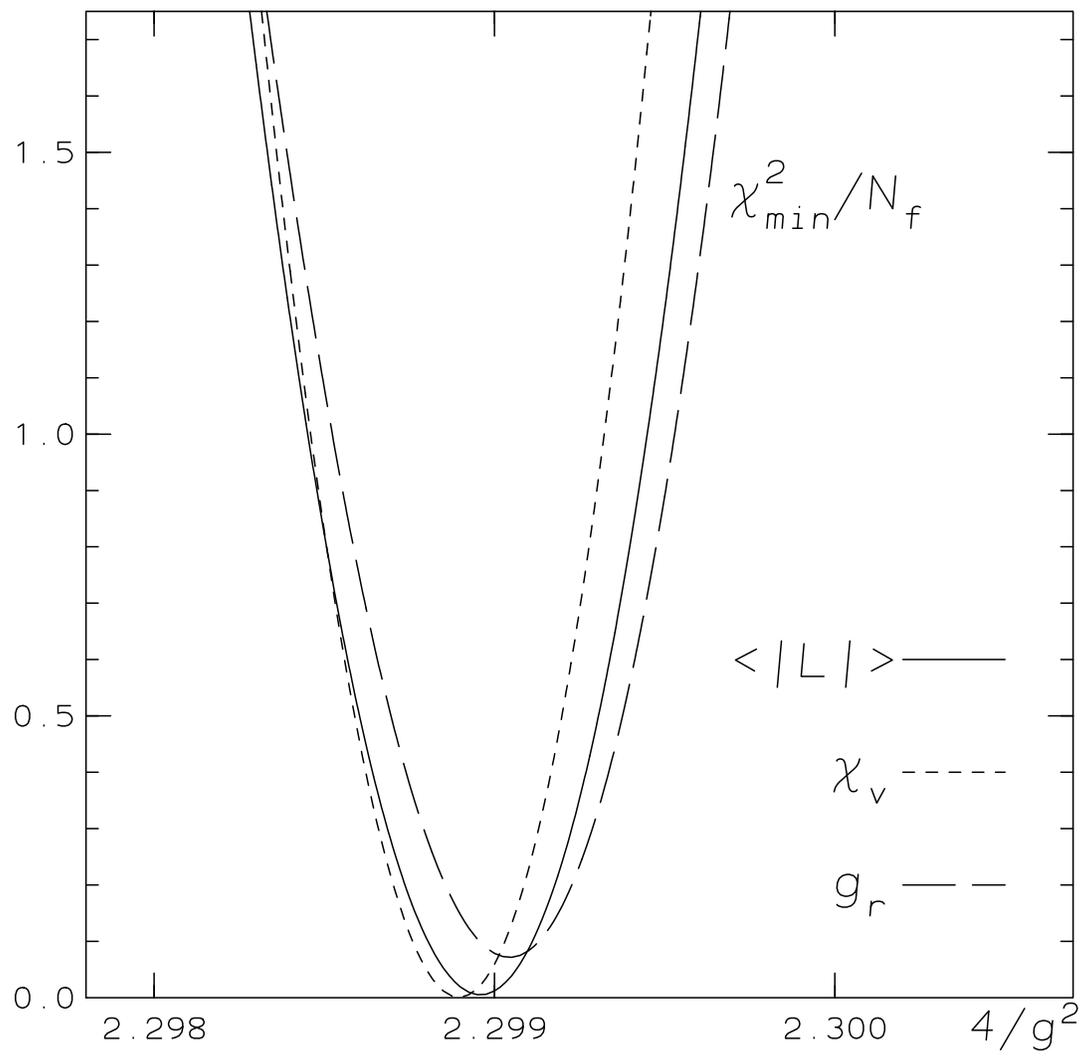
\item{Fig.~1}
The minimal values of $\chi^2$ per degree of freedom
for linear fits at fixed couplings $4/g^2$
according to \eq{lnex} for $\MO$ (solid line)
and $\chi_v$ (short dashes). For the cumulant $g_r$ (long dashes)
\eq{direx} with $\omega=1$ was used.
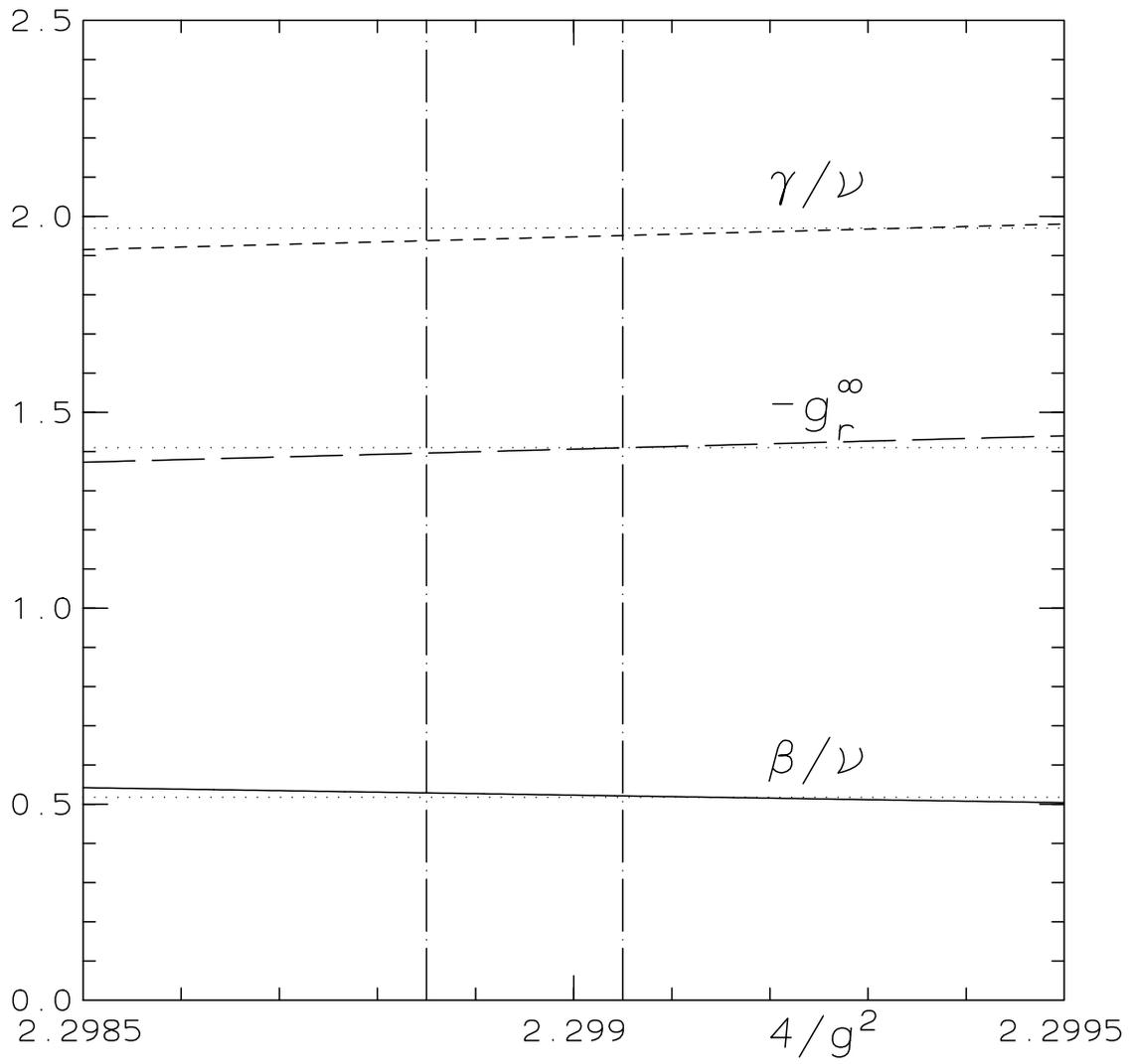
\item{Fig.~2}
The slopes $\beta/\nu$ (solid line) and $\gamma/\nu$ (short dashes),
and the constant $c_0=g_r^{\infty}$ (long dashes) from the same
fits as in Fig. 1. The dotted lines show the corresponding
$3d$ Ising model values, the dashed-dotted lines the error bars
of $4/g_{c,\infty}^2$.
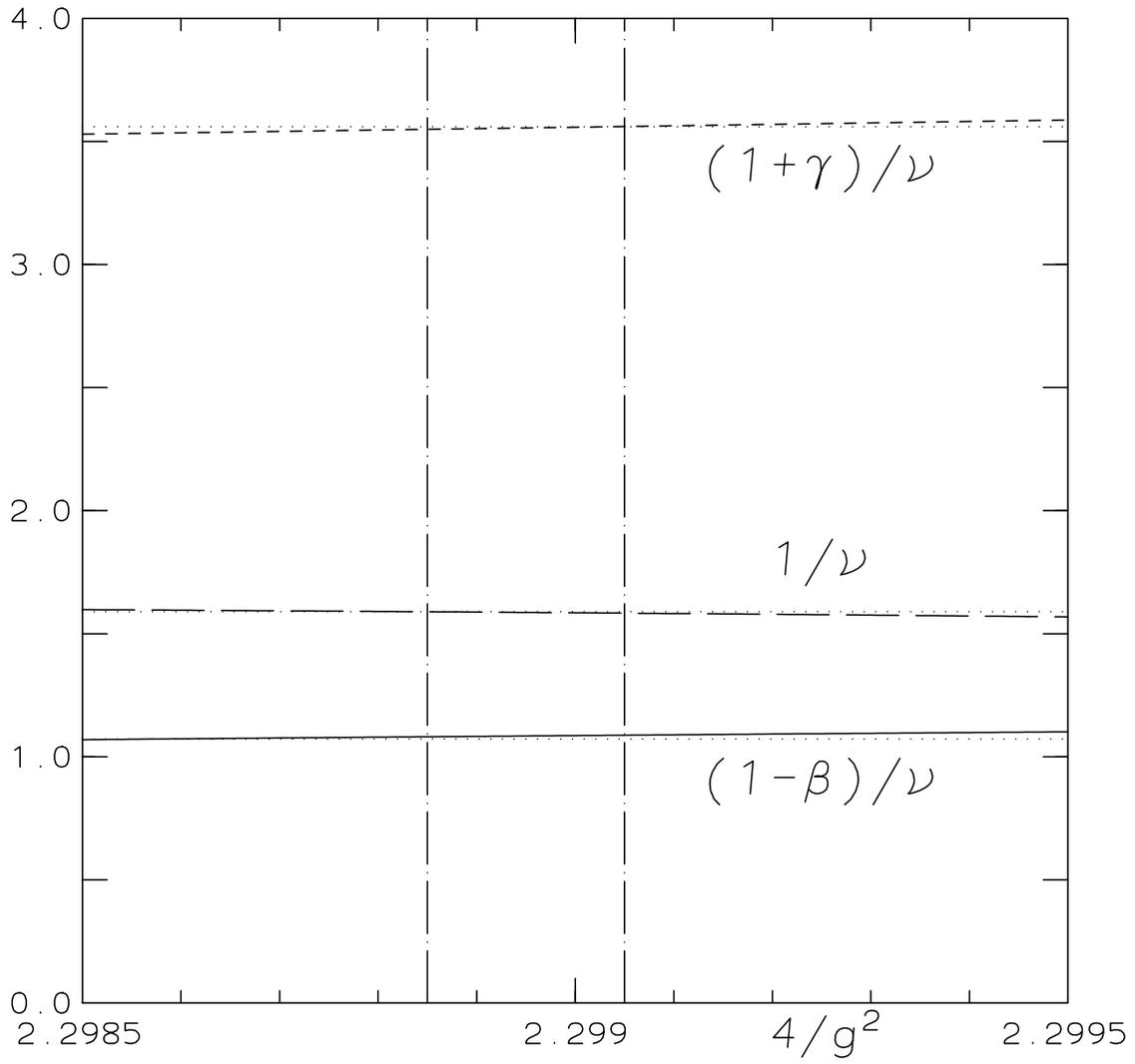
\item{Fig.~3}
The slopes of the linear fits to the logarithms of the derivatives
of $\MO$ (solid line), $\chi_v$ (short dashes)  and $g_r$ (long dashes).
The dotted lines show the corresponding $3d$ Ising model values,
the dashed-dotted lines the error bars of $4/g_{c,\infty}^2$.
\vfill\eject
\end